\documentstyle[aps,prl]{revtex}
\begin{document}

\draft
\title{{\tenrm\hfill JETP Lett. {\bf 61}, (1995)}\\
THE FOURTH-ORDER CORRELATION
FUNCTION OF \\ A RANDOMLY ADVECTED PASSIVE SCALAR}
\author{E. Balkovsky$^a$, M. Chertkov$^b$, I. Kolokolov$^c$,
and V. Lebedev$^{a,b}$}
\address{$^a$ Landau Inst. for Theor. Physics, Moscow, Kosygina 2,
117940, Russia \\
$^b$ Department of Physics of Complex Systems, Weizmann Inst. of Science,
Rehovot 76100, Israel\\
$^c$ Budker Inst. of Nuclear Physics, Novosibirsk 630090, Russia}
%\date{today}
\maketitle

\begin{abstract}
Advection of a passive scalar $\theta$ in $d=2$ by a large-scale velocity
field rapidly changing in time is considered. The Gaussian feature of the
passive scalar statistics in the convective interval was discovered in
\cite{95CFKLa}. Here we examine deviations from the Gaussianity: we obtain
analytically the simultaneous fourth-order correlation function of $\theta$.
Explicit expressions for fourth-order objects,
like $\langle(\theta_1-\theta_2)^4\rangle$ are derived.
\end{abstract}


\vskip 1cm

Advection of a passive scalar $\theta$ by an incompressible turbulent flow
is one of the classical problems in the theory of turbulence. The problem
is related to statistics of temperature or impurities in the flow.
The dynamics of the passive scalar is governed by
\begin{equation}
(\partial_t+u_{\alpha}\nabla_{\alpha}- \kappa
\triangle)\theta=\phi \,,
\label{eqm} \end{equation}
where the velocity ${\bf u}$ and the pumping $\phi$ are random functions of
$t$ and ${\bf r}$, $\kappa$ is the diffusion coefficient.  Correlation
functions of $\theta$ should be treated as averages both over statistics of
$\phi$ and {\bf u}. Batchelor \cite{59Bat} was the first who considered the
problem for a long-range velocity field. He found the pair correlator of the
passive scalar in the case where velocity field is very slow. Kraichnan
\cite{74Kra} considered the pair correlator in the opposite case of a
velocity field changing in time very rapidly. A theory for any finite
correlation time of the velocity field was proposed in \cite{95CFKLa}. It was
proved there that whatever be the statistics of the velocity field, the
statistics of the passive scalar in the convective interval approaches
Gaussianity as one increases the Peclet number (the ratio of the pumping
scale to the diffusion one).

In the present letter we aim at finding explicitly the fourth-order
correlation function of the passive scalar in Kraichnan regime.
We assume that the source $\phi$ is $\delta$-correlated in time and
spatially correlated on scale $L$:
\begin{equation}
\langle\phi(t_1,{\bf r}_1)\phi(t_2,{\bf r}_2)\rangle
=\delta(t_1-t_2)\chi_2(r_{12})\,,
\label{chi} \end{equation}
where $\chi_2(r)$ tends to zero at $r\gg L$.
We choose the simplest smooth form of the pumping
\begin{equation}
\chi_2(r)=P_2 L^2/(L^2+r^2) \,,
\label{pump} \end{equation}
where $P_2$ is the production rate of $\theta^2$.
A variation of the shape of $\xi_2(r)$ at $r\gtrsim L$
will keep all the values of interest to be intact at $r\ll L$.
The statistics of a velocity $\delta$-correlated in time is completely
defined by the pair correlation function which at dimension
$d=2$ equals
\begin{eqnarray} &&
\langle u_\alpha(t_1,{\bf r}_1) u_\beta(t_2,{\bf r}_2)\rangle=
\delta(t_1-t_2)\big[D L_u^2\delta_{\alpha\beta}-
{\cal K}_{\alpha\beta}({\bf r}_{12})\big] \,,
\label{large} \\ &&
{\cal K}_{\alpha\beta}(r)=D\big(
3\delta_{\alpha\beta}r^2/2-
r_\alpha r_\beta \big) \,,
\label{velo} \end{eqnarray}
where the incompressibility condition $\nabla\cdot{\bf u}=0$ is taken into
account. Here $D$ is a the characteristic strain describing the
strength of the velocity field, $L_u$ is the velocity correlation length
(the size of the largest vortex) which is assumed to be the largest scale in
the problem. The expression (\ref{velo}) is correct at $r\ll L_u$.

Simultaneous correlation functions of $\theta$ satisfy separate linear
equations \cite{Pri,94Kra,94SS}. The equation for the pair correlation
function $f(|{\bf r}_1-{\bf r}_2|)=\langle\theta_1\theta_2\rangle$
can be solved explicitly for arbitrary $r$. For separations $r\gg r_d$,
where $r_d=2\sqrt{\kappa/D}$ is the so-called diffusive length, one gets
\begin{equation}
f(r)=\frac{P_2}{2D}\left\{
\ln(1+L^2/r^2)+(L^2/r^2)\ln(1+r^2/L^2)\right\} \,.
\label{func} \end{equation}
The equation for the fourth-order simultaneous
correlation function of the passive scalar ${\cal F}=
\langle{\theta_1\theta_2\theta_3\theta_4}\rangle$ is
\begin{eqnarray} &&
-\hat{\cal L}{\cal F}({\bf r}_1,{\bf r}_2,{\bf r}_3,{\bf r}_4)
=\Phi(r_{12},r_{34})
+ \Phi(r_{13},r_{24})
+ \Phi(r_{14},r_{23}) \,,
\label{PP} \\ &&
\hat{\cal L}=\sum_{i>j}{\cal K}_{\alpha\beta}(r_{ij})
\nabla_{i\alpha}\nabla_{j\beta}
+\kappa \sum_{i=1}^4\triangle_i \,,
\label{op} \\ &&
\Phi(r_+,r_-)=
f(r_+)\chi_2(r_-)+f(r_-)\chi_2(r_+)\,,
\label{a2} \end{eqnarray}
where $r_{ij}=|{\bf r}_i-{\bf r}_j|$. The equation (\ref{PP}) can be
generalized for an arbitrary multi-scale velocity field and for an arbitrary
dimension $d$ \cite{94Kra,94SS,95CFKLb}. Although at any $d$ r.h.s. of
(\ref{PP}) decomposes into three parts each depending only on two
separations, only in the case of the large-scale velocity field the operator
$\hat{\cal L}$ supports the decomposed form: acting on an arbitrary function
of the two vectors it produces a function of the two vectors too. The
corollary gives us a possibility to establish the following general form of
the solution
\begin{equation}
{\cal F}({\bf r}_1,{\bf r}_2,{\bf r}_3,{\bf r}_4) =
F({\bf r}_{12},{\bf r}_{34})+
F({\bf r}_{13},{\bf r}_{24})+
F({\bf r}_{14},{\bf r}_{23})\,.
\label{a3} \end{equation}
Since we are looking for a solution in the restricted region $r\ll L_u$
one could add to (\ref{a3}) a zero mode of the operator $\hat{\cal L}$.
However, the equation (\ref{PP}) stems from the dynamical approach
(see \cite{95CFKLb} for details) that keeps no room for another solution
besides of the decomposed one. The function $F({\bf r}_+,{\bf r}_-)$ from
(\ref{a3}) satisfies the following equation
\begin{eqnarray} &&
-\bigl[\hat{\cal L}^\prime+\hat{\cal L}_d\bigr] F({\bf r}_+,{\bf r}_-)=
\Phi(r_+,r_-),
\label{a4} \\ &&
\hat{\cal L}^\prime=2D\sin^2\vartheta[\partial^2_y+
(\partial_\vartheta+\cot\vartheta\partial_\xi)^2],\quad
\hat{\cal L}_d=2\kappa
(\triangle_++\triangle_-)\,.
\label{a5} \end{eqnarray}
In (\ref{a5}) we passed to the variables $\xi=\ln[L^2/(r_+r_-)]$,
$y=\ln[r_-/r_+]$ and $\vartheta=\arccos[({\bf r}_+{\bf r}_-)/(r_+r_-)]$.
The physical boundary conditions imposed on $F$ are:
$F({\bf r}_+,{\bf r}_-)\rightarrow 0$ at $r_+$ or $r_-\to\infty$.

As was shown in \cite{95CFKLa}, correlators of $\theta$ do not feel
diffusion if $r_\pm\gg r_d$. Thus at $r\gg r_d$ we can omit the
diffusive term in (\ref{a4}) that together with the independence of
$\Phi(r_+,r_-)$ on $\vartheta$ allows us to write out a
solution of (\ref{a4}) as a sum $F=F_++F_-$,
where $F_\pm$ satisfy the equations
\begin{equation}
-2D \sin^2\vartheta\bigl[\partial_\vartheta^2+\partial_y^2\bigr]
F_\pm =f(r_\pm )\chi_2(r_\mp) \,.
\label{bbb} \end{equation}
Here $r_\pm=\sqrt{S/|\sin\vartheta|}\exp(\mp y/2)$ and $S,\vartheta,y$
should be treated as independent parameters. We conclude that,
by construction, $F_+(S,\vartheta,y)=F_-(S,\vartheta,-y)$.
Taking into account symmetry properties of (\ref{bbb}) we find that
$F_\pm$ is invariant under $\vartheta\to-\vartheta$ and under
$\vartheta\to\pi-\vartheta$. To solve the equation (\ref{bbb}) one can
use the resolvent ${\cal R}$ of the Laplacian
$\partial_\vartheta^2+\partial_y^2$ figuring in (\ref{bbb}):
\begin{equation}
-\big[\partial_\vartheta^2+\partial_y^2\big]
{\cal R}(\vartheta,\vartheta^\prime,y-y^\prime)=
\delta(\vartheta-\vartheta^\prime)\delta(y-y^\prime) \,.
\label{b014} \end{equation}
One should impose the zero boundary conditions on the resolvent at
the boundaries of the strip: $0<\vartheta<\pi$ and $-\infty<y<+\infty$,
since in accordance with the definition
$r_\pm\to\infty$ if $\vartheta\to 0,\pi$ or $y\to\pm\infty$.
The resolvent can be written explicitly:
\begin{equation}
{\cal R}=\frac{1}{4\pi}\ln\left[ \frac
{\sinh^2(y/2-y^\prime/2)+\sin^2(\vartheta/2+\vartheta^\prime/2)}
{\sinh^2(y/2-y^\prime/2)+\sin^2(\vartheta/2-\vartheta^\prime/2)}\right] \,.
\label{b015} \end{equation}
Convolution of the r.h.s. of (\ref{bbb}) with the resolvent gives
\begin{equation}
F_+=\frac{L^2P_2}{4D r_- ^2\sin^2\vartheta}
\int_0^\infty dw f(r_+ w)\tau^{-1}
\ln\biggl[\frac{(w_++\tau)^2+w^2\cot^2\vartheta}
{(w_-+\tau)^2+w^2\cot^2\vartheta}\biggr] \,,
\label{B21} \end{equation}
where $w_\pm=|w\pm w^{-1}|$,
$\tau=\sqrt{L^2r_-^{-2}\sin^{-2}\vartheta+w^{-2}}$.
Note that this formula leads to the weak angular singularity of $F_+$
at $\vartheta=0$:
\begin{equation}
\partial_\vartheta F_+\big|_{\vartheta=0}=
\pi P_2^2/D^2 \big(e^{2y} \ln(1+e^{-y})-e^y\big)\neq 0\,.
\label{fd} \end{equation}
Below we will show that the diffusion smoothes the singularity at the
smallest angles. Generally, (\ref{B21}) substituted into $F=F_++F_+(y\to-y)$
and further into (\ref{a3}) closes the problem of finding the fourth-order
correlation function ${\cal F}$ of the passive scalar for all separations
being larger $r_d$.

At $L\gg r_\pm$ the integration in the formula (\ref{B21}) can be performed,
then
\begin{eqnarray} &&
F=\frac{P_2^2}{D^2}\bigl\{
\ln^2[L/r_+]/2+\ln^2[L/r_-]/2+\ln[L/r_+] +\ln[L/r_-]+\pi^2/12 \bigr\}
\nonumber \\ &&
+{\cal Y}(z)+{\cal Y}(\overline{z})+{\cal O}(r_\pm/L),
\label{FPa} \\ &&
{\cal Y}(z)=\frac{P_2^2}{2D^2} \bigl\{\pi {\rm Im}\bigl[e^{-{\rm i}z}\bigr]
-{\rm Re} \bigl[\Omega(-{\rm i}e^{{\rm i}z})\bigr]
- {\rm Re}\bigl[ e^{-2{\rm i}z}\Omega({\rm i}e^{-{\rm i}z})\bigr] \,,
\nonumber \\ &&
\Omega(u)=\frac{\pi^2}{12} +\ln^2u
+{\rm Li}_2(1-{\rm i}u)+{\rm Li}_2(1+{\rm i}u)\bigr\} \,,
\label{FPb} \end{eqnarray}
where $z=\vartheta+{\rm i}y$ and $\overline{z}=\vartheta-{\rm i}y$,
and Li in (\ref{FPb}) is the integral logarithm.
Let us emphasize, that the second part of (\ref{FPa}),
$\propto[{\cal Y}(z)+{\cal Y}(\overline{z})]$, turns out to be
a zero mode of the Laplacian since it is the real part of an analytic
function of $z$. The general result (\ref{FPa}) shows a remarkable
coincidence of $F$ within a logarithmic accuracy with its collinear limit
$F(\vartheta=0)$. It follows from the dynamical approach described in
\cite{95CFKLa}. It was demonstrated there that $F$ is an average over
velocity's statistics of the product of times required for the vectors
${\bf r}_+$ and ${\bf r}_-$ respectively to grow upto $L$ if their edge
points move along Lagrangian trajectories. An evolution of the vectors, being
originally deep inside the convective interval $r_\pm\ll L$, can be divided
into two different stages: the first one is a quick collinearization of the
geometry, and second one is a long collinear stretching giving the major
contribution into the correlator. At $L\gg r_+\gg r_-$ (\ref{FPa}) gives
\begin{equation}
[F-f(r_+)f(r_-)]\approx \frac{P_2^2}{D^2}
\bigl\{\ln[L/r_+]+1+\vartheta^2-\pi|\vartheta|/2\bigr\} \,.
\label{as} \end{equation}
Here in the l.h.s. we subtracted from $F$ the major Gaussian part.

Let us now return to $F_+$ to examine its behavior at small angles.
The angular singularity (\ref{fd}) is formed by the $z$-dependent part of
(\ref{B21}) that is a zero mode of $\hat{\cal L}^\prime$. At small angles
it should be replaced by a zero mode ${\cal Z}$ of
$\hat{\cal L}^\prime+\hat{\cal L}_d$, where the diffusive operator
$\hat{\cal L}_d$ can be rewritten in terms of the set of variables
($\vartheta,y,\xi$) as
\begin{equation}
\hat{\cal L}_d=2\kappa \frac{e^\xi}{L^2} \cosh[y]
\bigl\{ \partial^2_\xi+\partial_y^2+
\partial_\vartheta^2+4\tanh[y]\partial_\xi \partial_y\bigr\}.
\label{dif} \end{equation}
The angular singularity is formed by the $z$-dependent part of
(\ref{B21}) which can be treated as a zero mode of $\hat{\cal L}^\prime$ by
analogy with (\ref{FPa}). At small angles it should be replaced by a zero
mode ${\cal Z}$ of the sum $\hat{\cal L}^\prime+\hat{\cal L}_d$.  We are
looking for a zero mode ${\cal Z}$ that is proportional to $|\vartheta|$ at
values of the angle, where it is possible to neglect the effect of diffusion
at all, and is regular at $\vartheta\to 0$. At the smallest value of
$\vartheta$ $\partial_\vartheta{\cal Z}$ is much larger than
$\partial_y{\cal Z}$ and $\partial_\xi{\cal Z}$. Thus, $Z$ should satisfy
\begin{equation}
\left\{(\vartheta^2\partial_\vartheta^2
+2\vartheta\partial_\vartheta \partial_\xi
-\partial_\xi +\partial_\xi^2)+A e^\xi
\partial_\vartheta^2\right\}{\cal Z}=0 \,,
\label{ja} \end{equation}
where $A=\kappa\cosh(y)/(16 D L^2)$ can be considered as an arbitrary
parameter. Taking into account (\ref{fd}) we find the following solution
of (\ref{ja}) possessing the desirable behavior
\begin{equation}
{\cal Z}\approx\pi \frac{P_2^2}{D^2}
\big[e^{2y}\ln(1+e^{-y})-e^{y}\big]
\sqrt{\vartheta^2+\kappa\cosh y e^\xi/(DL^2)} \,.
\label{jg} \end{equation}
The characteristic angle
$\vartheta_0=\sqrt {\kappa D^{-1}L^{-2}\cosh[y]} \exp(\xi/2)$
turns out to be small at $r_\pm\gg r_d$. Using the explicit form (\ref{jg})
we conclude that for $\vartheta\sim \vartheta_0$:
$\partial_\vartheta{\cal Z}\sim \vartheta_0^{-1}{\cal Z}$,
$\partial_y{\cal Z}\sim{\cal Z}$, $\partial_\xi{\cal Z}\sim{\cal Z}$.
Those estimations justify above calculations. To summarize, at $r_\pm\gg r_d$
the diffusion is relevant only at small angles $\vartheta\lesssim
\vartheta_0$ where it influences angular derivatives of $F$ but it gives a
negligible correction to the expression (\ref{B21}).

As long as one considers a correlation function at sufficiently
small distances, say $F$ at $r_-\lesssim r_d$, the account of diffusion is
unavoidable. The major value of the function can be found directly from the
suitable expression in the convective interval by putting there,
in terms formally divergent at $r\to 0$, $r_d$ instead of the smallest
distances (see Appendix A of \cite{95CFKLa} for the proof). Thus, using
(\ref{FPa}), we find with the logarithmic accuracy the following fourth-order
objects
\begin{eqnarray} &&
\mbox{at}\quad L\gg r_{13},r_{14},r_{23},r_{24}\gg
r_{12}>r_{34}\gg r_d,\quad
{\bf r}_{12} \, || \, {\bf r}_{34}
\nonumber \\ &&
\langle (\theta_1-\theta_2)^4\rangle\approx
12 \frac{P_2^2}{D^2}\bigl(\ln^2[r_{12}/r_d]+\ln[r_{12}/r_d]\bigr) \,,
\label{corra} \\ &&
\langle (\theta_1-\theta_2)^2\theta_3^2\rangle-
\langle (\theta_1-\theta_2)^2\rangle \langle\theta_3^2\rangle\approx
2\frac{P_2^2}{D^2}\ln[r_{12}/r_d]\,,
\label{corrb} \\ &&
\langle (\theta_1-\theta_2)^2(\theta_3-\theta_4)^2\rangle-
\langle (\theta_1-\theta_2)^2\rangle \langle(\theta_3-\theta_4)^2\rangle
\approx 4\frac{P_ 2^2}{D^2}\ln[r_{34}/r_d] \,.
\label{corrc} \end{eqnarray}
(\ref{corra}) shows that with the logarithmic accuracy
($f(0)\approx f(r_d)$) $\langle (\theta_1-\theta_2)^2\rangle = 2f(0)-2f(r_{12})
\approx 2P_2/D \ln[r_{12}/r_d]$. Thus, it follows from (\ref{corra}) that the
flatness $\langle (\theta_1-\theta_2)^4\rangle/ \big( \langle
(\theta_1-\theta_2)^2\rangle \big)^2 \approx 3$ in accordance with the
asymptotic Gaussianity proved in \cite{95CFKLa}. To overcome from the
correlators (\ref{corrb},\ref{corrc}) to respective correlators of the
dissipative field $\epsilon=\kappa(\nabla\theta)^2$ we should differentiate
them over $r_{12},r_{34}$ and replace the separations after all by $r_d$. One
gets the following estimations
\begin{eqnarray}
\mbox{at}\quad L\gg r_{13}\gg r_d \quad &&
\langle\langle \epsilon_1\theta_3^2\rangle\rangle=\kappa
\langle\langle (\nabla\theta_1)^2\theta_3^2\rangle\rangle
\sim{P_2^2}/D  \,,
\label{gra} \\ &&
\langle\langle \epsilon_1\epsilon_3\rangle\rangle=\kappa^2
\langle\langle (\nabla\theta_1)^2(\nabla\theta_3)^2
\rangle\rangle\sim P_2^2 \,.
\label{grad} \end{eqnarray}
Unknown multipliers behind the parametric dependencies in the formulas
(\ref{gra},\ref{grad}) are of the order of unity. We postpone an explicit
calculation of the multipliers, that requires a direct account of diffusion.
The formulas (\ref{gra},\ref{grad}) show (in accordance
with \cite{95CFKLb}) the zero dimensionality both for the passive scalar
$\theta\sim r^0$ and the dissipation field $\epsilon\sim r^0$.

Note that in the region
$L_u\gg r_{ij},r_{kl}\gg L,r_{ik},r_{il},r_{jk},r_{jl}$
the found correlator ${\cal F}$ does not decay. The decay occurs only at the
largest scales $r_{12},r_{34}\gg L_u$ which are out of the scope of the
present study.

The proposed solution for the fourth-order correlation function of $\theta$
could serve a starting point for studying passive scalar correlations in
the general case of a multi-scale (but short-correlated in time) velocity
field, the problem is introduced by Kraichnan \cite{68Kra}.
A closure \cite{94Kra,95KYC}, applied recently to the model, yields an
anomalous scaling, particularly, for $\langle(\theta_1-\theta_2)^4\rangle$ that
conflicts with an analytic consideration \cite{95CFKLb}.
Future attempts of a perturbative study based on the results of the present
letter could solve the collision. From another hand, an inclusion of
the present results into the scheme proposed in \cite{94FL,94FLa} could help
in better understanding the problem of the direct cascade in two-dimensional
turbulence.

We thank G. Falkovich for numerous valuable discussions and
R. Kraichnan for useful comments.
V. L. is grateful for the support of the Minerva Center
for Nonlinear Physics.

\end{document}